\documentclass[a4paper,12pt]{article}
\usepackage{epsfig}\usepackage{amssymb}
\newcommand\FIGURE[5]{\begin{figure}[htb]\centering\epsfig{file=#1.ps,
            width=#2\linewidth,angle=#3}\caption{#4}\label{#5}\end{figure}}
\newcommand\FIGGER[3]{\begin{figure}[htb]\centering\epsfig{file=#1.ps,
            width=#2\linewidth,angle=#3}\end{figure}}
\title{A Dynamical System with Two Strange Attractors.}
\author{R. C. Johnson\thanks{{\tt bob.johnson@dur.ac.uk}}
 \\ {\em Department of Mathematical Sciences},\\ {\em University of Durham},
    {\em Durham DH1 3LE}, {\em England}.}
\begin{document}
\maketitle
\begin{abstract}A six-dimensional R\"ossler-Lorenz hybrid
has two coexistent attractors.  Both, either or neither may be
strange.\end{abstract}
\paragraph{Introduction.} 
The R\"ossler \cite{rossler} and Lorenz \cite{lorenz} equations are textbook
examples of chaotic dynamical systems. Each is three-dimensional,
continuous-time, smooth and autonomous, with a single strange attractor for
certain parameter values.

This paper draws attention to a six-dimensional hybrid with two coexistent
attractors, one or both of which may be strange.

The attractors evidently stem from the constituent R\"ossler and Lorenz
systems, but each requires and occupies the larger phase space.

The new system may well have potential for physical modelling, but its
discovery is the result of curiosity.
\paragraph{Construction.} 
Start with the Lorenz equations \cite{lorenz} 
\begin{eqnarray}\dot x&=&\sigma(y-x)\nonumber\\ 
\dot y&=&(r-z)x-y\label{eLRNZ}\\ \dot z&=&xy-\beta z\nonumber\end{eqnarray} 
in a chaotic regime with 
\begin{equation}\sigma=10,\quad r=28,\quad\beta=\textstyle{8\over3}.
\label{eLPAR}\end{equation} 
Fig.~1(a) shows a corresponding solution trajectory on the familiar
\cite[p.~283]{geometry} two-winged strange attractor.
	
Consider replacing $z$ on the right-hand side in the last component by some
external excitation $w(t)$: 
\begin{equation}xy-\beta z\quad\to\quad xy-\beta w.\label{eZ}\end{equation} 
The response may be regular or irregular, depending on $w$.

For instance, with $w=\cos t$, there is steady oscillation about values
$$x=y=0,\quad z=r=28.$$If instead the excitation $w(t)$ is an isolated
pulse then the response is a sudden excursion of $z(t)$ followed later by
sudden relaxation --- each change accompanied by spikes of $x(t)$ and $y(t)$.

Now, the chaotic R\"ossler system \cite{rossler}
\begin{eqnarray}\dot u&=&-v-w\nonumber\\
	\dot v&=&u+av\label{eRSLR}\\
	\dot w&=&b+(u-c)w\nonumber\end{eqnarray}
with parameters\begin{equation}a=0.2,\quad b=0.2,\quad
                                     c=5.7\label{eRPAR}\end{equation}
has solution trajectories on a well-known \cite[p.~287]{geometry}
twisted-band strange attractor --- as Fig.~1(b) shows. 
	
Evidently its third component $w(t)$ provides an irregular sequence of
isolated pulses. Such a sequence, via eq.~(\ref{eZ}), provokes the Lorenz
$(x,y,z)$ variables into ragged oscillation about $(0,0,r)$.

This chaotic ringing, however, does not have composite properties. For
delay-coordinate reconstruction \cite{ntsa} from $z(t)$ reveals a
R\"ossler-like twisted-band attractor in three embedding dimensions. Lorenz
adds nothing.

Higher-dimensional and richer structure is possible --- but needs feedback
from Lorenz to R\"ossler. And one amoung many options is to mix the
oscillations of the $x$--$y$ pair into those of $u$ and $v$.

Experiments eventually lead to linear driving of $v$ by $x$ and $y$ --- with
further linear coupling from $u$ to $x$ --- as follows:
\begin{eqnarray}\dot u&=&-v-w\nonumber\\ 		
                \dot v&=&x+ay\nonumber\\ 		
                \dot w&=&b+(u-c)w\nonumber\\
		\dot x&=&\sigma(y-x)+\mu u\label{eROLO}\\
		\dot y&=&(r-z)x-y\nonumber\\
		\dot z&=&xy-\beta w.\nonumber\end{eqnarray}
This is the system of interest. With other parameters fixed, adjustments of
$\mu$ and $\sigma$ give one or two strange attractors, or a cycle plus a
strange attractor, or two cycles, or a single cycle.
\paragraph{Properties.}
If parameters $r$, $\sigma$, $\beta$, $a$, $b$, $c$ take the values in
eqs.~(\ref{eLPAR}) and (\ref{eRPAR}), phase-space volumes shrink rapidly.

For $\sigma=10$, the system of eq.~(\ref{eROLO}) has two coexisting
attractors if $\mu$ is between about $0$ and $13$. They have the following
properties.\begin{itemize}
\item
The two attractors centre approximately around $u=x=y=0$ and $z=r=28$ but
are displaced in $v$ and $w$. See Figs.~2 and 3. Rectangular phase-space
boxes just sufficient to enclose them differ in volume by a factor of
typically $10^5$.
\item
The larger attractor exists for $\mu\lesssim13$, where it is always strange.
Its toadstool-like $xyz$ projection --- Fig.~2(a) --- is reminiscent of the
Lorenz attractor in Fig.~1(a). For this reason it is labelled $L$.
\item
The smaller one exists for $\mu\gtrsim0$ and is strange for
$\mu\lesssim3.26$ --- see Fig.~3. As $\mu$ increases beyond $3.26$ (through
50 and beyond) it alternates between regular --- i.e., a cycle --- and
strange, bifurcating by period-doubling. This parallels the behaviour of the
Lorenz attractor as $r$ increases \cite{sparrow}.
\item
Its twisted-band $uvw$ projection --- Fig.~3(b) --- persists with changing
$\mu$. Because it is similar to the R\"ossler attractor in Fig.~1(b) this
attractor is labelled $R$.
\item
The two attractors have $v_L<v_R$ and $w_L>w_R$ and, where they coexist,
trajectories move much faster round $L$ than around $R$ --- see Fig.~4.
\item
Fig.~4 also shows more variability of period and amplitude on $R$ than
on $L$, and trajectories on $R$ and $L$ with $\mu=1$ are found to have
maximum Lyapunov exponent of about $1.4$ and $0.45$ respectively.
\item
Both attractors, where they are strange, are higher-dimensional objects. For
delay-coordinate reconstruction \cite{ntsa} of $R$ and $L$ from samples of
components $u,v,\dots,z$ shows that they need about 5 embedding dimensions.
\end{itemize}
These properties depend in different ways on $r$, $\beta$, $a$, $b$, $c$
and $\sigma$.
\begin{itemize}\item
A change of $r$, for instance, shifts $R$ and $L$ bodily along the
$z$-axis. And while minor variations in $a$, $b$, $c$ and $\beta$ affect
other numerical details, there are larger changes with $\sigma$.
\item With $\mu=1$, $L$ is a cycle for $2.8\lesssim\sigma\lesssim4.5$ ---
and then $R$ is also a cycle for $\sigma\approx3.7$ --- see Fig.~5.
\item As Fig.~5(a) suggests, both cycles bifurcate to strange attractors by
period-doubling. In this regime they overlap in all but their $v$ component
--- they still have $v_L<v_R$.
\end{itemize}
Note that when $R$ and $L$ vanish with parameter changes it is because their
basins of attraction shrink. No examples have been found where they
coalesce.

Thorough exploration of six-dimensional phase space, varying seven
parameters, is a formidable task. However, investigation of the
$\mu$-$\sigma$ plane is promising, and some further illustrations are
available elsewhere \cite{www}.
\paragraph{Conclusion.}
As it stands, the system of eq.~(\ref{eROLO}) is no more than an interesting
construction without direct physical origin. It is just one in a range of
possible R\"ossler-Lorenz hybrids.

Its immediate value is to show that six-dimensional phase space is roomy
enough to accommodate two attractors not connected by (e.g.) symmetry ---
and that all combinations of chaotic and regular oscillation may coexist.

Dynamical models that use in the order of six effective dimensions occur
in for instance physiology (e.g. \cite{phys}), meteorology (e.g.
\cite{weath}) and economics (e.g. \cite{econ}) --- where the possibility of
such coexistent modes could be very interesting.
\paragraph{Acknowledgement.} The trajectory calculations used {\sl XPPAUT}
\cite{bard}. Lyapunov exponents and embedding dimension were determined
with {\sl NDT} \cite{ndt}. The authors are applauded for the functionality
and availability of their software.
%
  
%
\newpage
\section*{Figure captions}
\begin{description}
\item[Figure 1](a) Solution of the Lorenz system eq.~(\ref{eLRNZ}) with
	parameters as in eq.~(\ref{eLPAR}) --- a trajectory on its
	two-winged strange attractor. 
	(b) Solution of the R\"ossler system eq.~(\ref{eRSLR}) with 
	parameters as in eq.~(\ref{eRPAR}) --- a trajectory on its
	twisted-band strange attractor.
\item[Figure 2]System of eq.~(\ref{eROLO}) at $\mu=1$ with other
	parameters as in eqs.~(\ref{eLPAR}) and (\ref{eRPAR}). The
	$L$-attractor --- (a) $xyz$-projection of a trajectory; 
	(b) $uvw$-projection of a trajectory.
\item[Figure 3]System of eq.~(\ref{eROLO}) at $\mu=1$ with other
	parameters as in eqs.~(\ref{eLPAR}) and (\ref{eRPAR}). The
	$R$-attractor --- (a) $xyz$-projection of a trajectory;
	(b) $uvw$-projection of a trajectory.
\item[Figure 4]System of eq.~(\ref{eROLO}) at $\mu=1$ with
	other parameters as in eqs.~(\ref{eLPAR}) and (\ref{eRPAR}).
	Component $z(t)$ of trajectories on (a) the $L$-attractor and
	(b) the $R$-attractor. Compare the time scales and
	variability of period and amplitude.
\item[Figure 5]System of eq.~(\ref{eROLO}) at $\mu=1$ with $\sigma=3.7$
	and otherwise parameters as in eqs.~(\ref{eLPAR}) and (\ref{eRPAR}).
	(a) The $L$-attractor --- $xyz$-projection of a trajectory. 
	(b) The $R$-attractor --- $uvw$-projection of a trajectory.
	Both attractors are cycles, and $L$ is doubled.
\end{description}
%
\newpage
\FIGGER{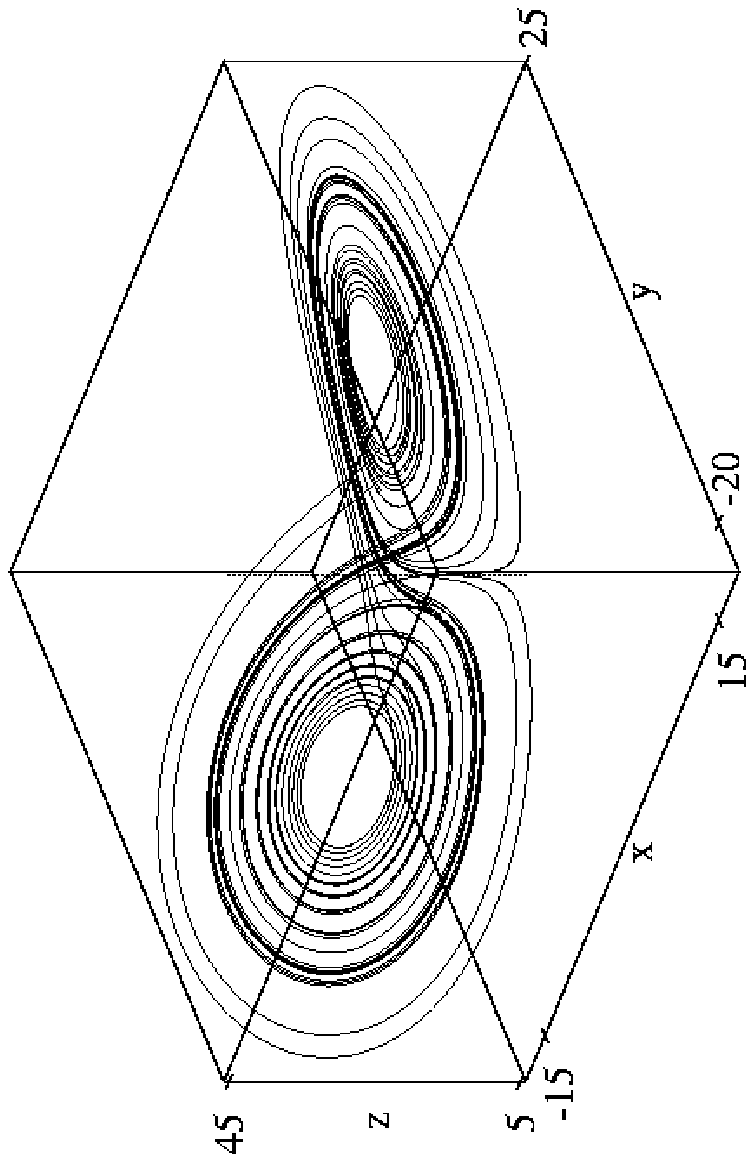}{.5}{-90}
\FIGURE{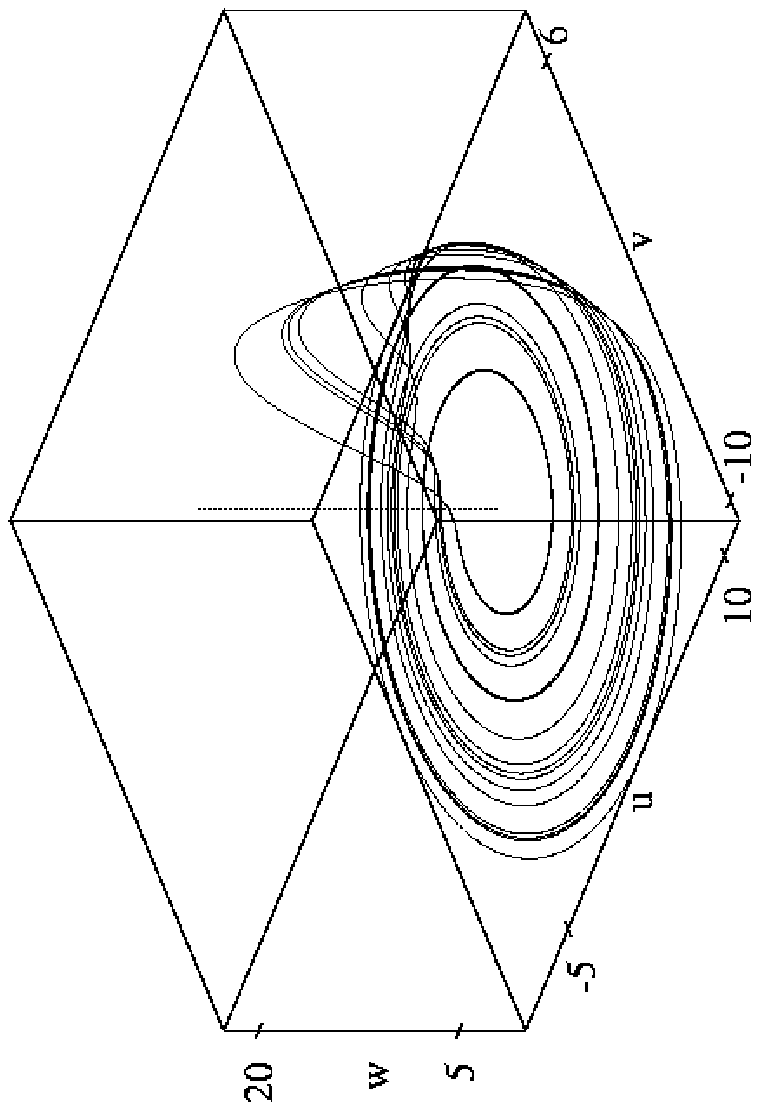}{.5}{-90}{(a) above, (b) below}{fONE}
\FIGGER{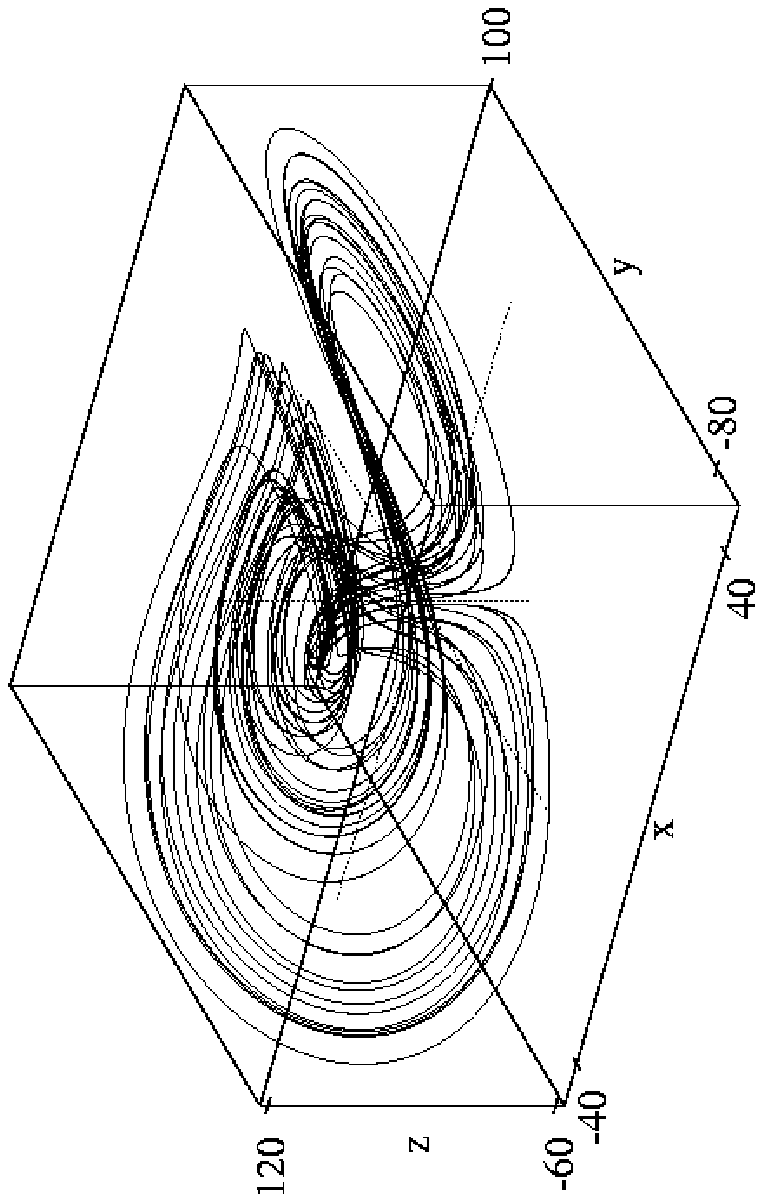}{.5}{-90}
\FIGURE{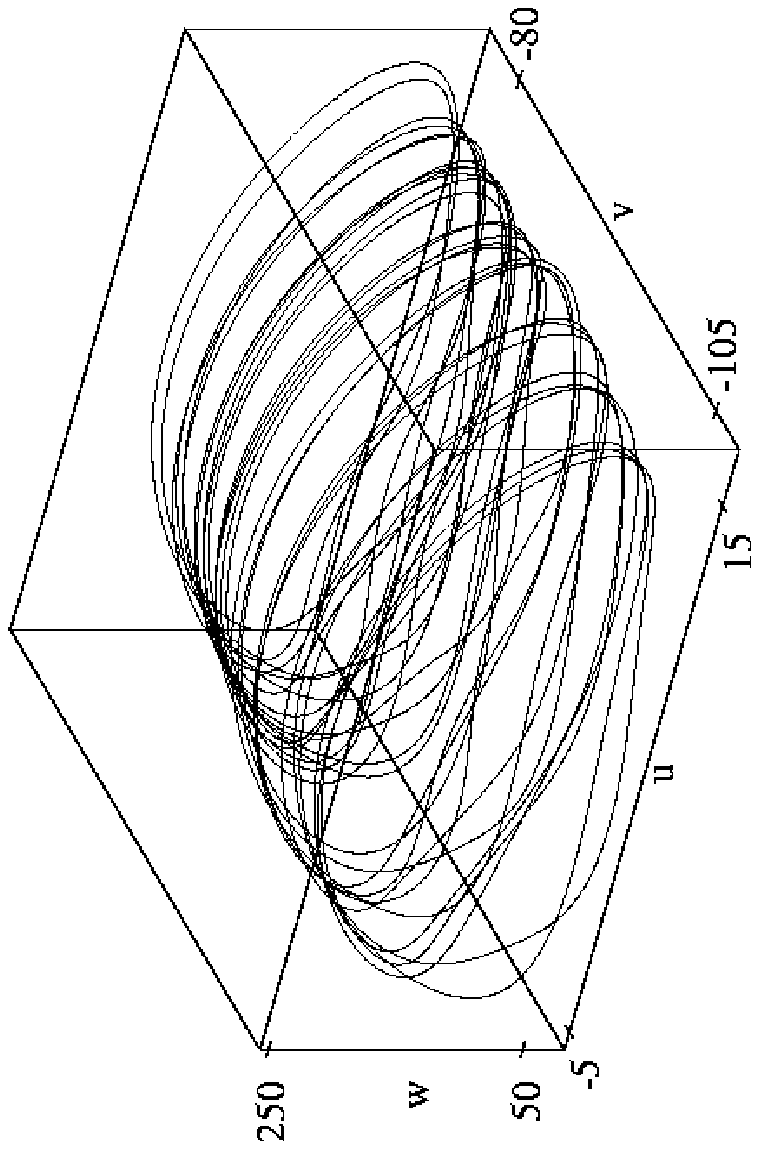}{.5}{-90}{(a) above, (b) below}{fTWO}
\FIGGER{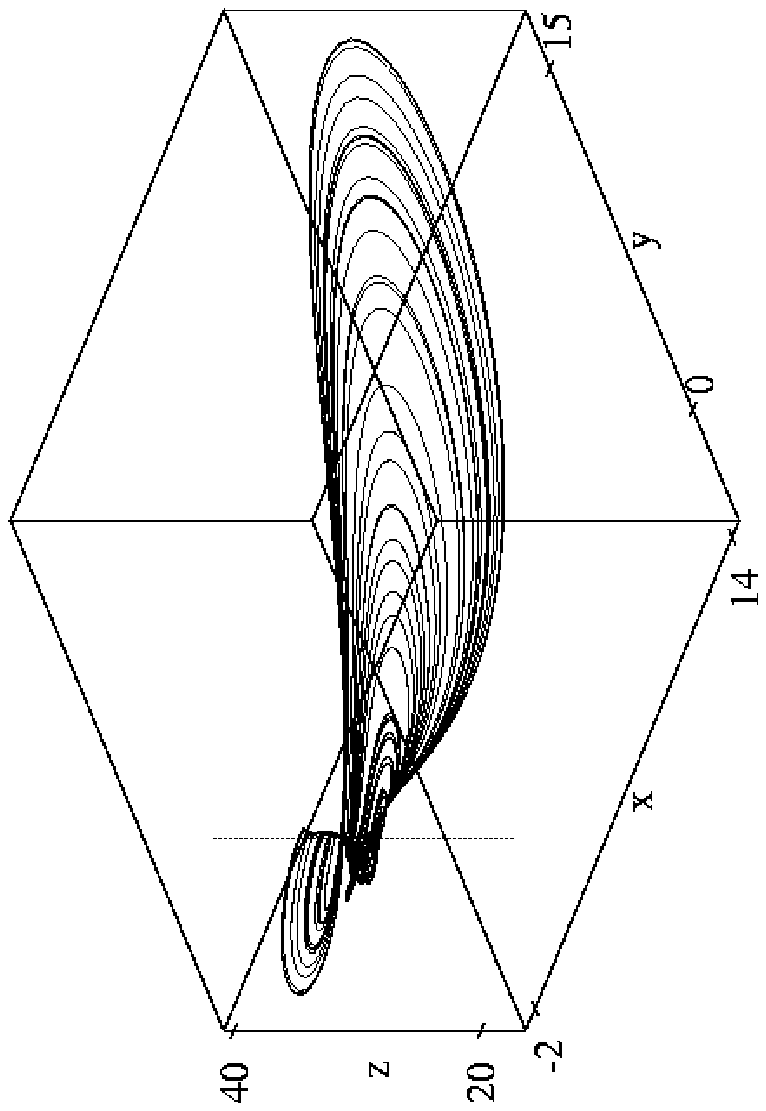}{.5}{-90}
\FIGURE{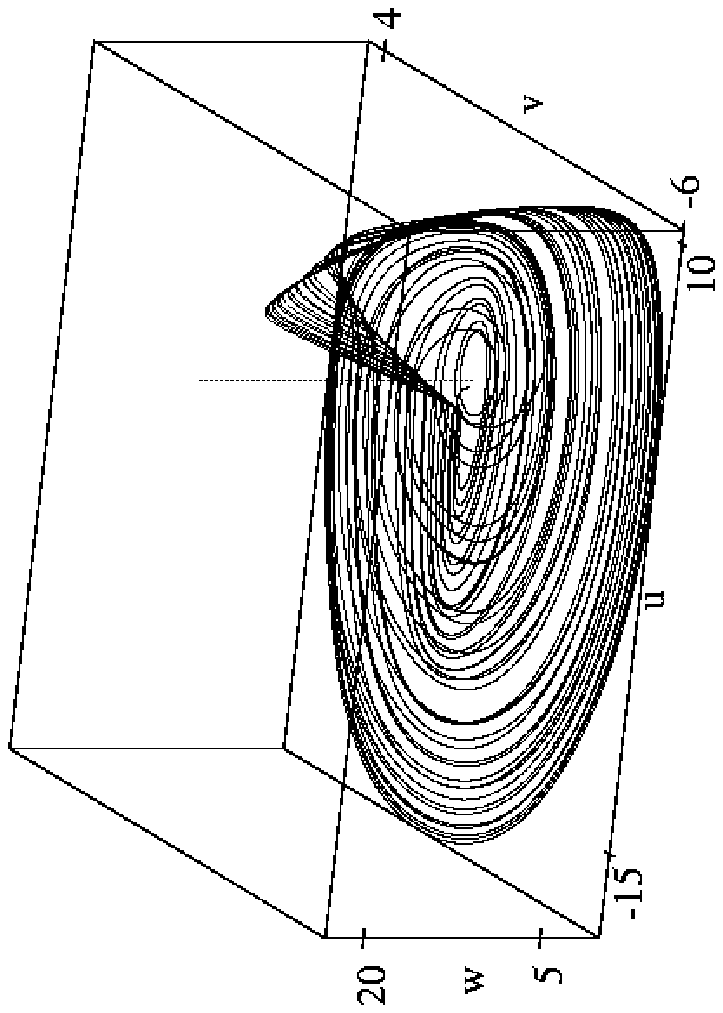}{.5}{-90}{(a) above, (b) below}{fTHREE}
\FIGGER{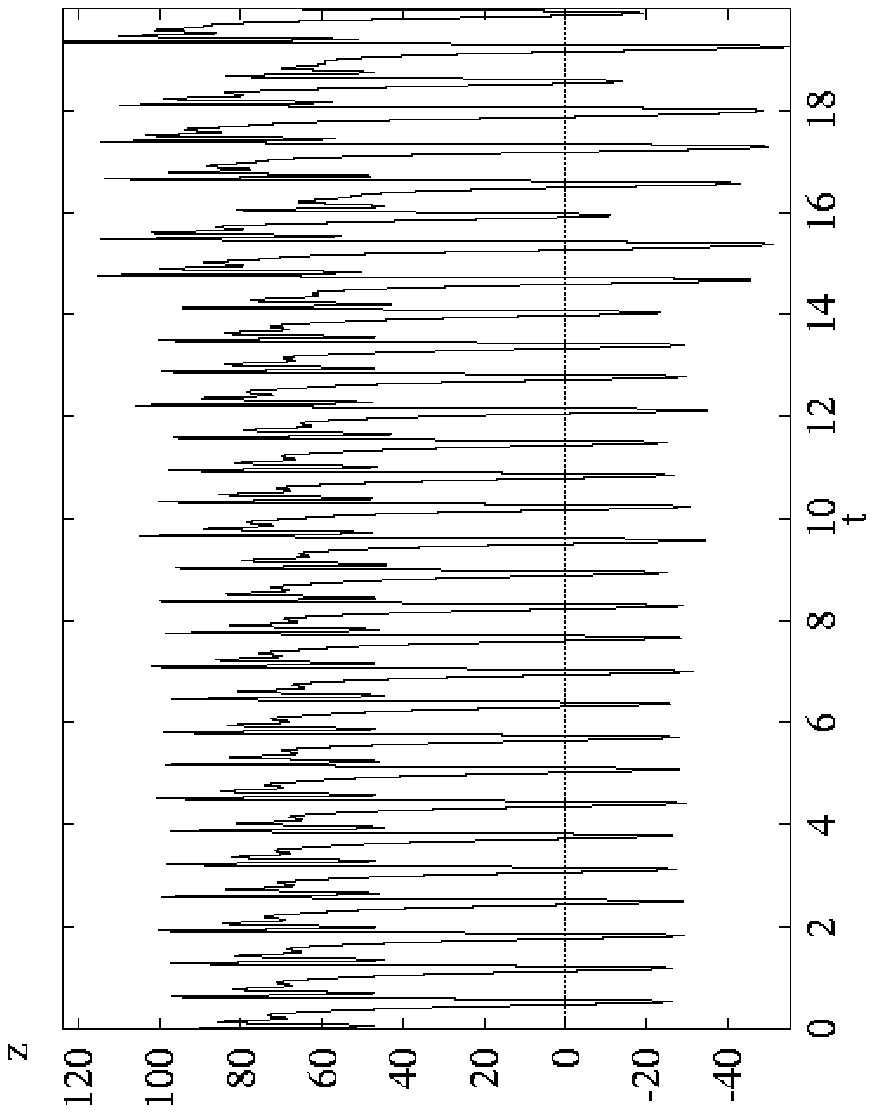}{.5}{-90}
\FIGURE{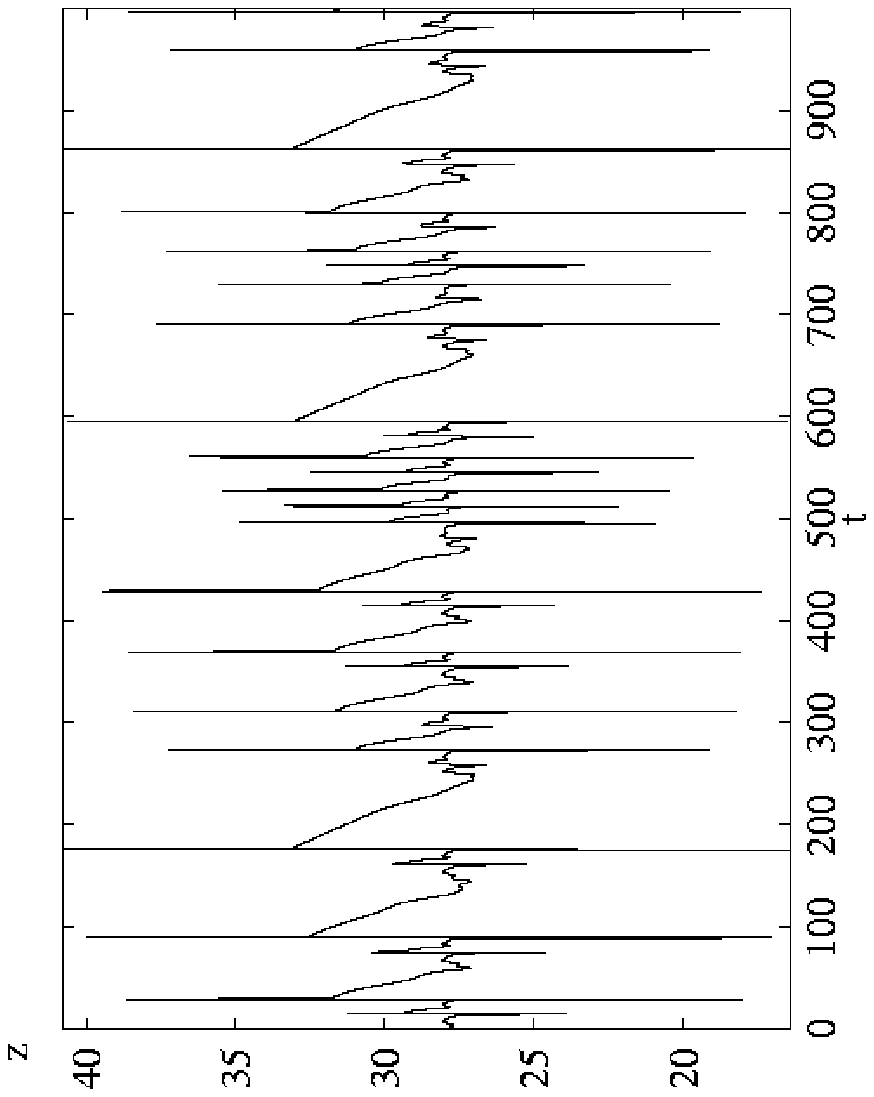}{.5}{-90}{(a) above, (b) below}{fFOUR}
%
\FIGGER{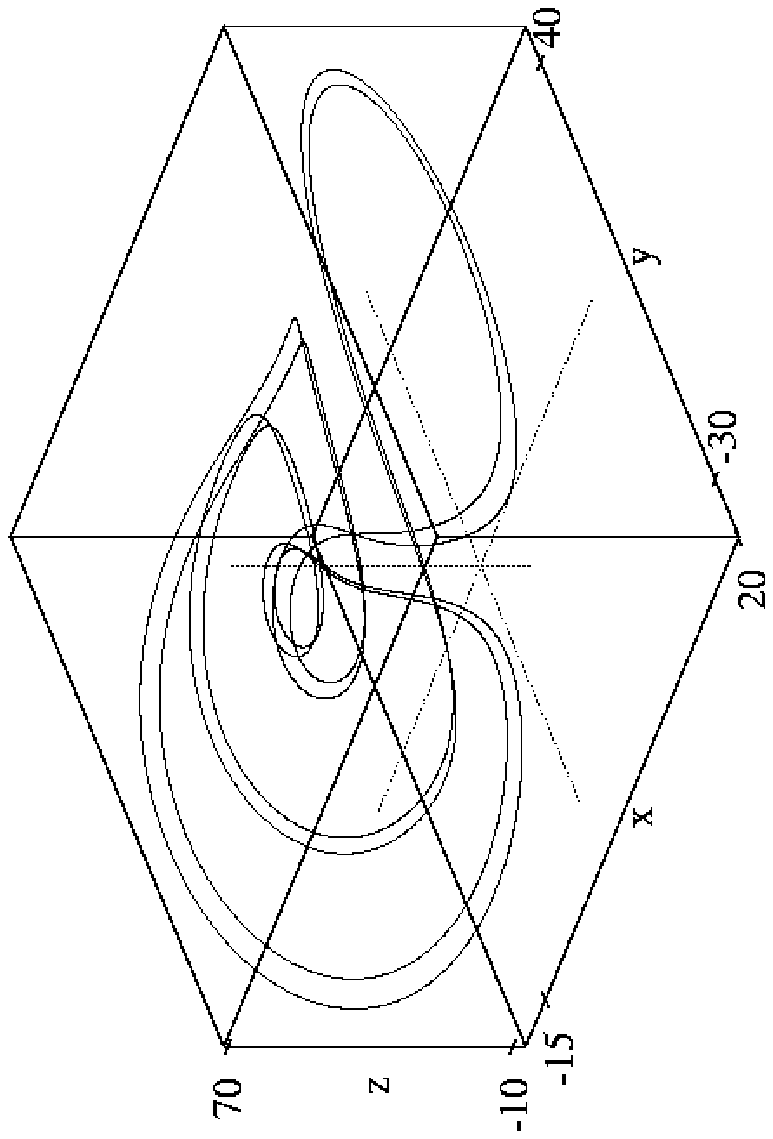}{.5}{-90}
\FIGURE{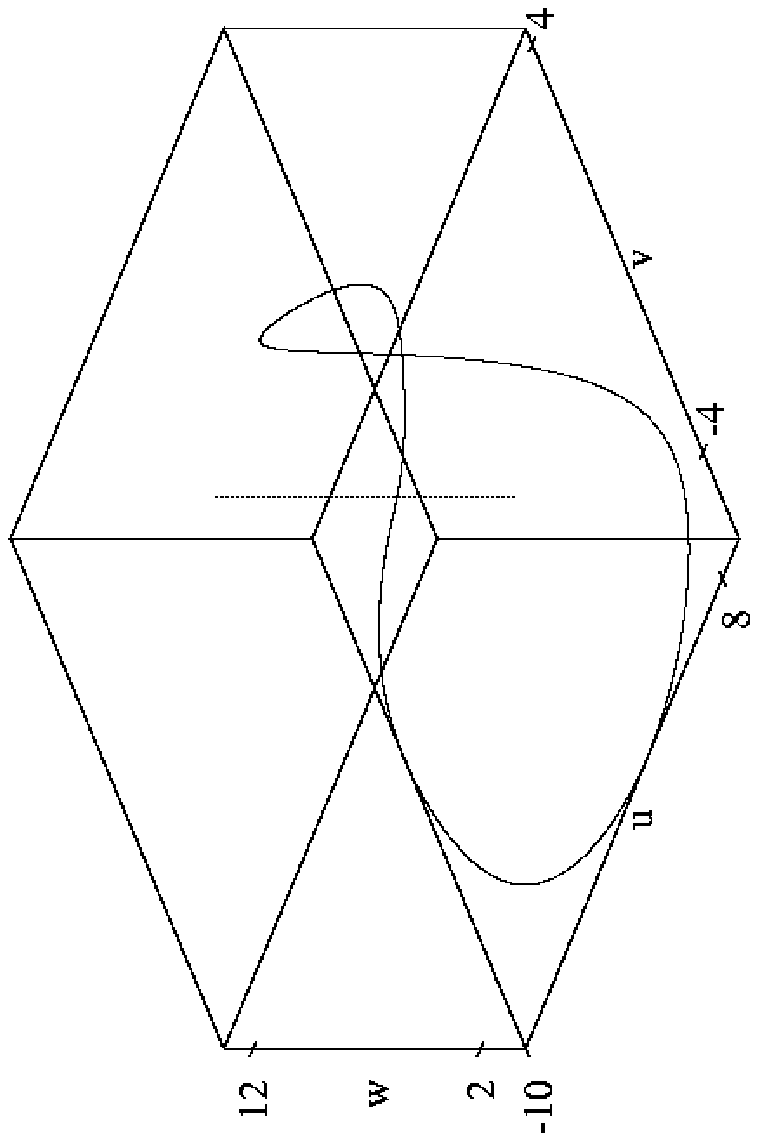}{.5}{-90}{(a) above, (b) below}{fFIVE}
\end{document}